\documentclass{iau307}
\usepackage{graphicx}
\usepackage{natbib}
\usepackage{url}
\usepackage{dtklogos}
\bibpunct{(}{)}{;}{a}{}{,}

\newcommand{\beq}{\begin{equation}}
\newcommand{\eeq}{\end{equation}}
\newcommand{\beqa}{\begin{eqnarray}}
\newcommand{\eeqa}{\end{eqnarray}}

\newcommand{\nbeq}{\begin{equation*}}
\newcommand{\neeq}{\end{equation*}}

\title[magnets as probes] %% give here short title %%
{Constraining general massive-star physics by exploring the unique properties
of magnetic O-stars: \ Rotation, macroturbulence \& sub-surface convection}

\author[Sundqvist]   %% give here short author list %%
{Jon O. Sundqvist$^1$}
\affiliation{$^1$Universit\"atssternwarte, Scheinerstr. 1, 81679
M\"unchen, Germany \\ email: {\tt mail@jonsundqvist.com}} %\\[\affilskip]
\pubyear{2014}
\volume{307} 
\pagerange{}
\jname{New windows on massive stars: asteroseismology, interferometry, and spectropolarimetry}
\editors{G. Meynet, C. Georgy, J.H. Groh \& Ph. Stee, eds.}
\begin{document}

\maketitle

\begin{abstract}

A quite remarkable aspect of non-interacting O-stars with detected
surface magnetic fields is that they all are very slow rotators.  This
paper uses this unique property to first demonstrate that the
projected rotational speeds of massive, hot stars, as derived using
current standard spectroscopic techniques, can be severely
overestimated when significant ``macroturbulent'' line-broadening is
present. This may, for example, have consequences for deriving the
statistical distribution of rotation rates in massive-star
populations, and for the use of these rates in stellar evolution
models. It is next shown how such macroturbulence (seemingly a
universal feature of hot, massive stars) is present in all but one of
the magnetic O-stars, namely NGC\,1624-2. Assuming then a simple model
in which NGC\,1624-2's exceptionally strong, large-scale magnetic
field suppresses atmospheric motions down to layers where the magnetic
and gas pressures are comparable, first empirical constraints on the
formation depth of this enigmatic hot-star macroturbulence are
derived. The results suggest an origin in the thin sub-surface
convection zone of massive stars, consistent with a physical origin
due to, e.g., stellar pulsations excited by the convective motions.

\end{abstract}

\firstsection % if your document starts with a section,
\section{Introduction} 

Over the past decade, new generations of spectropolarimeters and large
survey programs have revealed that roughly $\sim$\,10\,\% of all
massive main-sequence stars harbor large-scale, organized surface
magnetic fields, quite similar to those of intermediate-mass ApBp
stars (see, e.g., \citealt{Wade12}; Grunhut, this Volume). The fields
are strong, typically on the order of kG, and their fundamental origin
is basically unknown, although recent observations of Herbig pre-main
sequence stars point toward surviving fossils from early phases of
stellar formation \citep{Alecian13}. A particularly neat property of
these magnetic massive stars is that they are \textit{oblique
  rotators} (meaning their magnetic and rotation axes are offset), so
that their rotation periods can be readily measured from the observed
variation of the line-of-sight field \citep[e.g.,][]{Borra80} or from
photometric/spectral variations caused by their circumstellar
magnetospheres \citep[e.g.,][]{Howarth07}. This paper focuses on
(non-interacting) magnetic O-stars, which all have very long measured
rotation periods (likely because they have been spun down through
magnetic braking by their strong stellar winds,
e.g. \citealt{Petit13}). By means of high-quality spectra collected
within the Magnetism in Massive Stars project (MiMeS,
\citealt{Wade12}), I use these unique properties to examine:

\begin{itemize}
	\item The accuracy of standard methods for inferring rotation
          rates of massive stars.
	\item General origin (and magnetic inhibition) of
          "macroturbulence" in hot stars.
\end{itemize} 

\section{Rotation and macroturbulence in massive, hot stars} 

For most stars, it is not possible to directly measure the rotation
rate. Instead one typically infers the \textit{projected} stellar
rotation, $v \sin i$ (with inclination angle $i$), from observed,
broadened line-spectra. However, it is since long known that rotation
is not the only macroscopic broadening agent operating in hot star
atmospheres. The additional broadening is of very large width,
typically on order $\sim$\,50\,km/s (well in excess of the
photospheric speed of sound, $\sim$\,20\,km/s), and the occurrence of
this ``macroturbulence'' seriously complicates deriving accurate $v
\sin i$ rates for massive stars that are not too rapidly rotating
\citep[e.g.,][]{Howarth97, SimonDiaz14}. Moreover, since early-type
stars lack surface convection associated with hydrogen recombination
-- which is responsible for such non-thermal broadening in late-type
stellar atmospheres \citep{Asplund00} -- the physical origin of
macroturbulence in hot stars remains unclear (though see, e.g.,
\citealt{Aerts09}). At the present, it is normally treated by simply
introducing ad-hoc photospheric velocity fields with Gaussian
distributions of speeds, assumed to be either \textit{isotropic} or
directed only \textit{radially and/or tangentially} to the stellar
surface.

\textbf{Properties of the magnetic O-stars.} Table.~\ref{Tab:params}
summarizes relevant parameters for the sample of magnetic O-stars
considered here, including a non-magnetic comparison star
(HD\,36861). The table includes derived values of characteristic
(isotropic) macroturbulent velocities $\theta_{\rm G}$ from
\citet{Sundqvist13}, obtained by using information about $v \sin i$
from the measured rotation periods. Note in particular two things from
this table: i) The long rotation periods of the magnetic stars indeed
imply $v \sin i \approx 0$\,km/s for several of them, and ii) strong
macroturbulent line-broadening is present in all but one of the
magnetic O-stars, namely NGC\,1624-2.

%%Since the rotational line-broadening thus unambigously is negligible
%%for most of these stars, they are perfect objects for testing standard
%%methods for inferring $v \sin i$ in cases where the rotation period is
%%not a-priori known; the next section performs such a test.
%%Furthermore, the second point directly suggests there is something
%%different with NGC\,1624-4, as compared to the other stars in the
%%sample. Sect. 4 examines this in more detail, showing how it can give
%%important constraints on the general origin of macroturbulence in
%%massive, hot stars.

\begin{table}
\begin{minipage}{\textwidth}
    \centering
    \caption{Stellar and magnetic parameters for the sample O-stars,
      including $v \sin i$ as implied from the measured rotation
      periods and macroturbulent velocities $\theta$ (assuming here
      isotropic macroturbulence, $\theta_{\rm G}$, see text). Table
      adapted from \citet{Sundqvist13}.}
        \begin{tabular}{ l l l l l l l l }\\
        \hline \hline Star & Spec. type &$T_{\rm eff}$ & $\log g$     & $B_{\rm pole}$ &$P_\mathrm{rot}$ & $v \sin i$ & $\theta_{\rm G}$ \\
                                 && [\,kK\,] & [\,cgs\,]                 & [\,kG\,] &[d]& [\,km\,s$^{-1}$\,] & [\,km\,s$^{-1}$\,] \\ \hline
                NGC\,1624-2    & O6.5-O8\,f?cp & 35 & 4.0 & 20     &158& 0 & $2.2\ \ \pm\,^{0.9}_{2.2}$ \\ 
                HD\,191612         & O6\,f?p-O8\,f?cp & 35 & 3.5 & 2.5     &538& 0 & $62.0\,\pm\,^{0.5}_{0.5}$ \\
                 HD\,57682             & O9\,V & 34 & 4.0 & 1.7     &64& 0 & $19.2\,\pm\,^{0.3}_{0.3}$ \\
                CPD\,-28\,2561     & O6.5\,f?p & 35 & 4.0 & 1.7     &70& 0 & $24.3\,\pm\,^{1.0}_{0.9}$ \\
                HD\,37022         & O7\,Vp & 39 & 4.1 & 1.1     &15& 24 & $42.9\,\pm\,^{0.5}_{0.6}$ \\
                HD\,148937         & O6\,f?p & 41 & 4.0 & 1.0     &7& 45 & $54.0\,\pm\,^{0.9}_{0.9}$ \\
                HD\,108             & O8\,f?p & 35 & 3.5 & 0.5     &1.8\,$\times 10^4$& 0 & $64.4\,\pm\,^{0.4}_{0.4}$ \\
                HD\,36861         & O8\,III((f)) & 35 & 3.7 & 0     &--& 45 & $50.0\,\pm\,^{0.3}_{0.3}$ \\
                \hline
        \end{tabular}
    \label{Tab:params}
\end{minipage}
\end{table}

\section{Do standard methods overestimate $v \sin i$?}

\begin{figure} 
% \begin{minipage}{6.5cm}
 \begin{minipage}{2.8cm}
%   \hspace{1.3cm}
%   \centering  
%   \vspace{-3cm}
   \resizebox{\hsize}{!}
%    {\includegraphics[width=7cm,angle=90]{Fig1a.eps}}
    {\includegraphics[angle=90]{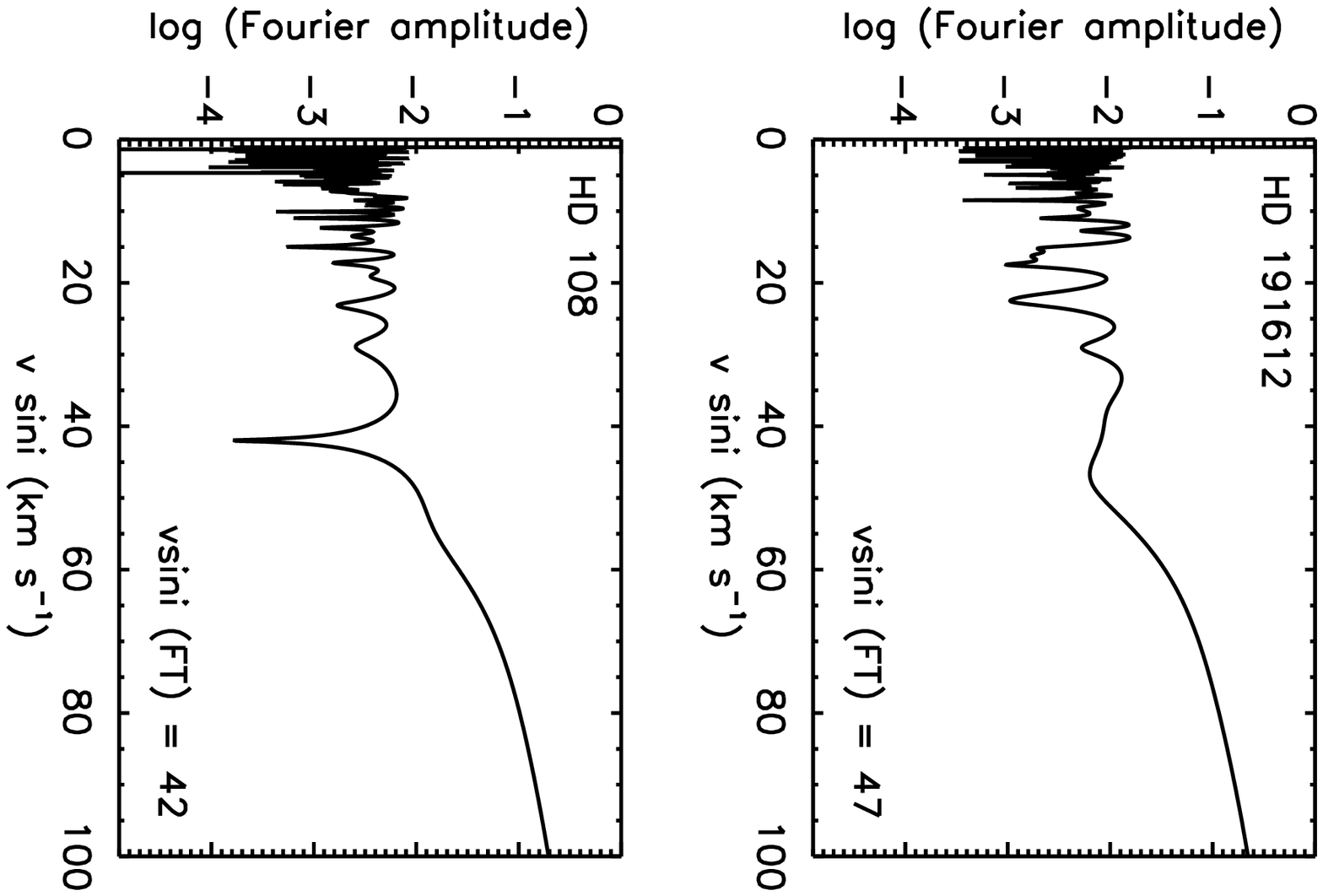}}
 \end{minipage}
 %
% \begin{minipage}{6.5cm}
 \begin{minipage}{5.5cm}
   \centering  
   \resizebox{\hsize}{!}
%    {\includegraphics[width=5.0cm,angle=90]{Fig1b.eps}}
    {\includegraphics[angle=90]{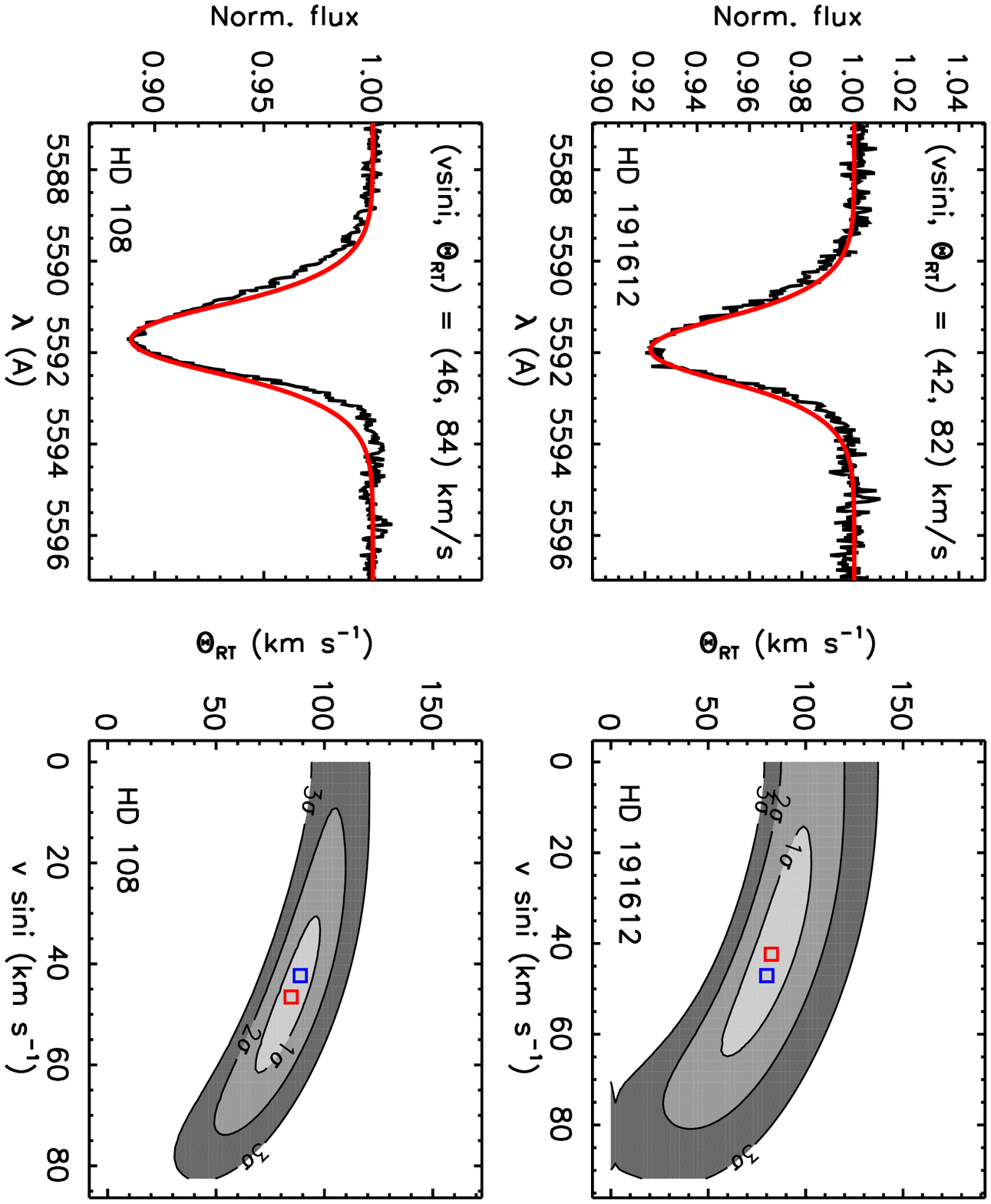}}
 \end{minipage}
 \begin{minipage}{5.5cm}
   \centering  
   \resizebox{\hsize}{!}
%    {\includegraphics[width=5.0cm,angle=90]{Fig1c.eps}}
    {\includegraphics[angle=90]{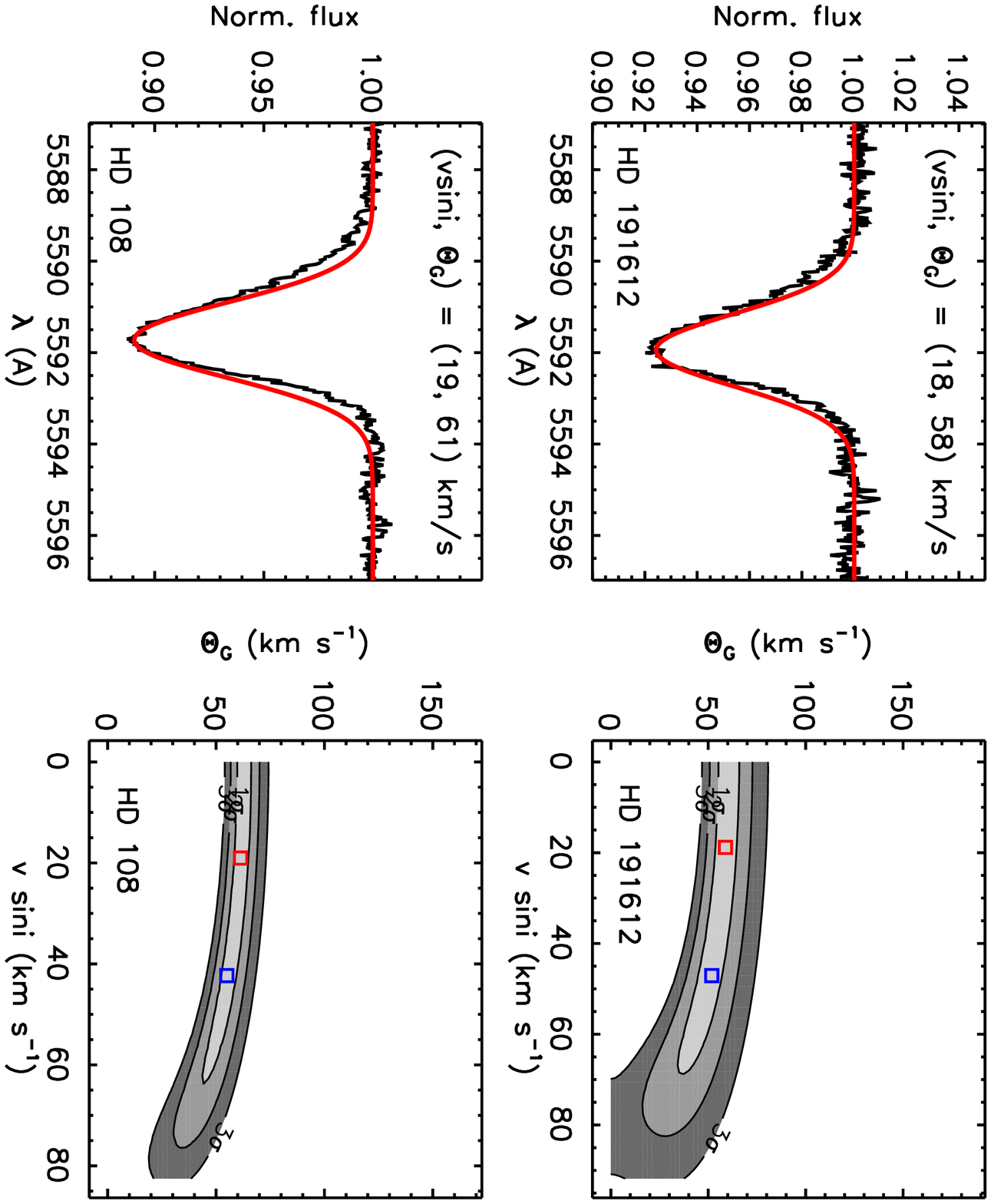}}
 \end{minipage}

\caption{Projected rotation speeds $v \sin i$ and macroturbulent
  velocities $\theta$ for HD\,191612 and HD\,108, derived using
  standard FT (left panel) and GOF (middle/right panels)
  techniques. The contour-maps show 1,2,3$\sigma$ confidence intervals
  for the fits in the $v \sin i$-$\theta$ plane. The blue and red
  squares on the contour-maps indicate the FT derived value and the
  best GOF model, respectively. The middle panel assumes a
  radial-tangential macroturbulence with equal contributions from both
  directions, $\theta_{\rm RT}$, and the right panel assumes isotropic
  velocity fields, $\theta_{\rm G}$. The true values for $v \sin i$
  are $<$ 1 km/s for both stars, see text. Adapted from
  \citet{Sundqvist13b}.}
\label{Fig:vsini} 

\end{figure}

I here follow \citet{Sundqvist13b} and use HD\,191612 and HD\,108 as
test-beds, the two stars in Table~\ref{Tab:params} with $v \sin i < 1$
km/s and characteristic macroturbulent velocities $\theta_{\rm G} >
50$\,km/s.
%%Of these two stars, HD\,191612 has the shorter rotation period,
%%$P = 538$ days, which along with its estimated stellar radius $14.5
%%R_\odot$ and inclination angle $i \approx 35-50^{\circ}$ yields $v
%%\sin i < 1$ km/s \citep{Howarth07, Sundqvist12b}. For the even longer
%%period of HD\,108, $v \sin i < 1$ km/s independent of $i$ \citep[see,
%%  e.g.,][]{Petit13}. 
Fig.~\ref{Fig:vsini} shows the results from deriving $v \sin i$ and
macroturbulent velocities for HD\,191612 and HD\,108, using the
standard Fourier Transform (FT) and Goodness-of-fit (GOF)
techniques. As illustrated by the figure, the FT method derives $v
\sin i$ from the position of the first minimum in Fourier space,
whereas the GOF method convolves synthetic line-profiles for a range
of $v \sin i$ and macroturbulent velocities, creating a standard
$\chi^2$-landscape from which a best-combination of the two parameters
is determined (see \citealt{SimonDiaz14} for
details). Fig.~\ref{Fig:vsini} illustrates how the FT method yields $v
\sin i \approx 40-50$ km/s for both stars, a severe overestimate
compared to the true value $v \sin i < 1$ km/s. The best GOF model
assuming radial-tangential macroturbulence also gives $v \sin i
\approx 40-50$ km/s, whereas assuming isotropic macroturbulence
actually results in lower $v \sin i \approx 20$ km/s, although then of
course the results from the FT and GOF methods do not
agree\footnote{Note also that the derived characteristic velocities are
  quite different depending on which form of macroturbulence is
  assumed, due to the markedly different shapes of a disc-integrated
  radial-tangential velocity field model and an isotropic one.}.
Agreement in the derived $v \sin i$ between these two methods has
indeed been used as an argument in favor of the radial-tangential
macroturbulence model \citep[e.g.,][]{SimonDiaz14}, but the analysis
here shows clearly that such agreement does not necessarily mean the
derived $v \sin i$ is correct. The GOF contour-maps in
Fig.~\ref{Fig:vsini} further display quite wide ranges of allowed
values of $v \sin i$. Particularly for isotropic macroturbulence the
results are degenerate all the way down to zero rotation, rendering
the ``best'' model from this GOF quite useless (in contrast to the
well-constrained values of $\theta_{\rm G}$ in Table 1, derived using
independent knowledge of $v \sin i$).

Overall, these results illustrate a severe problem regarding deriving
$v \sin i$ in the presence of a macroturbulent broadening that
significantly influences the appearance of the line profile. In the
case here of slow rotators, blindly applying standard methods leads to
drastic overestimates of $v \sin i$, where the results also depend on
the assumptions made about the unknown velocity fields causing the
additional broadening. The next section now shows how we may indeed
use the magnetic O-stars to also shed some light on the physical
origin of this enigmatic macroturbulence.

\section{Constraining the origin of macroturbulence by exploring magnetic inhibition of hot-star sub-surface convection}

\begin{figure} 
 \begin{minipage}{5.0cm}
   \hspace{0.5cm}
%   \centering  
%   \vspace{-3cm}
   \resizebox{\hsize}{!}
    {\includegraphics[]{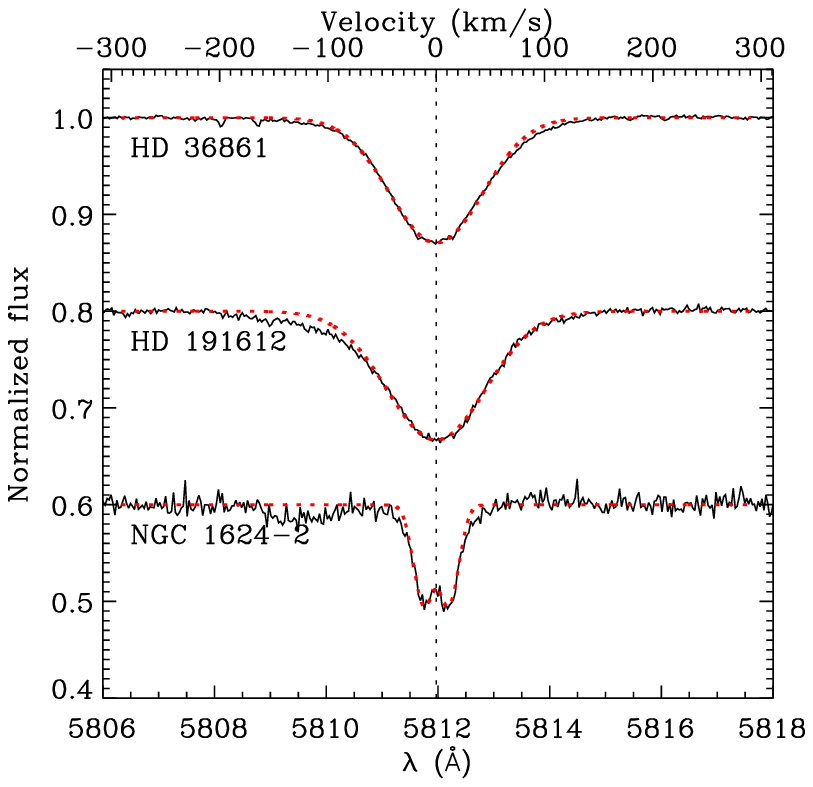}}
 \end{minipage}
 \begin{minipage}{6.0cm}
   \hspace{1.0cm}
%   \centering  
   \vspace{-0.3cm}
   \resizebox{\hsize}{!}
    {\includegraphics[angle=90]{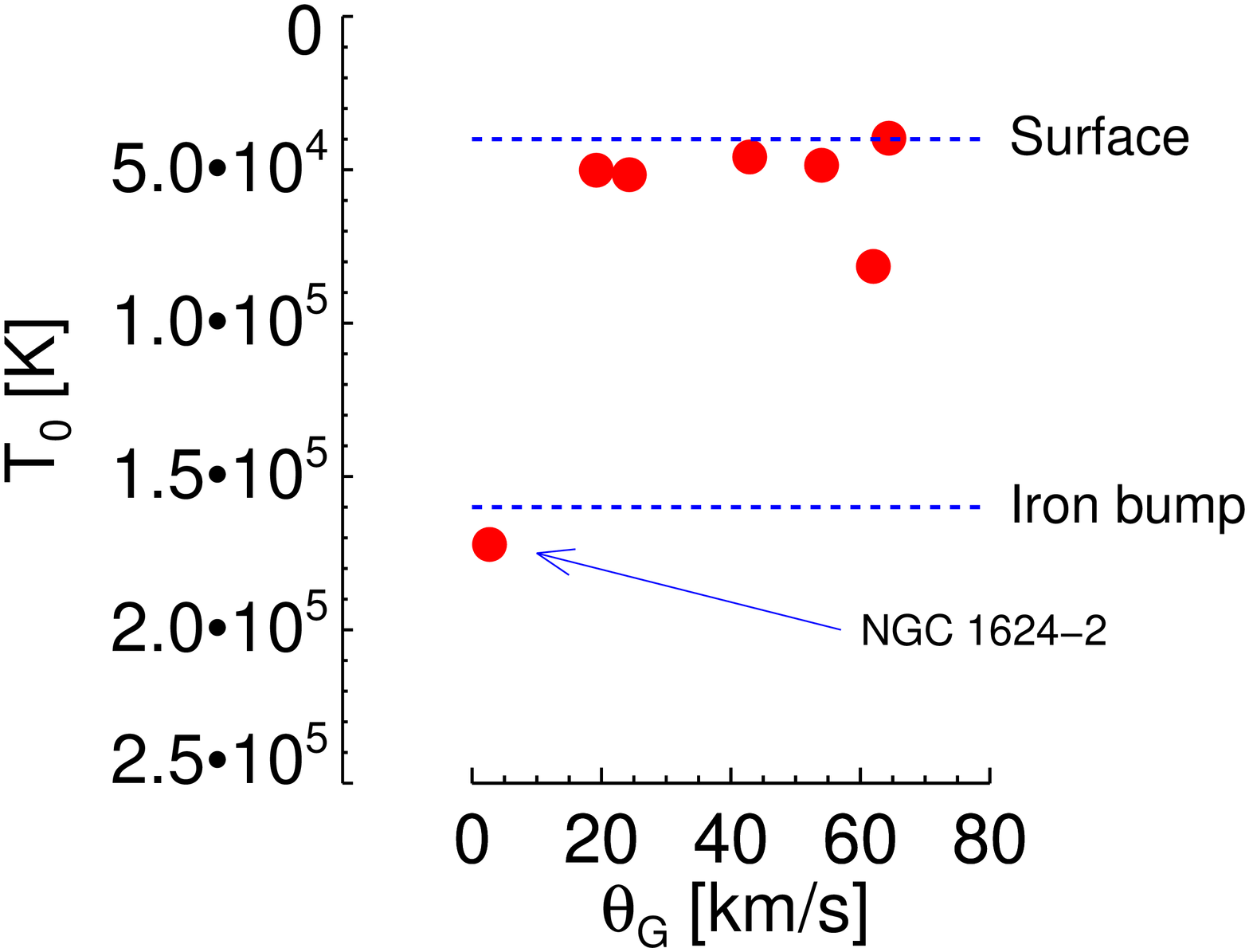}}
 \end{minipage}
 \caption{\textbf{Left panel:} Observed (black solid) and fitted (red
   dashed) {\sc c\,iv} $\lambda\lambda$5812 line profiles for three
   stars in our sample, as labeled in the figure. The horizontal
   dashed line marks line center, and the continua in the two lower
   curves have been shifted downwards by 0.2 and 0.4 normalized flux
   units. \textbf{Right panel:} Atmospheric temperature $T_0$ at
   which $\beta = 1$ (see text) vs. macroturbulent velocity
   $\theta_{\rm G}$ for the magnetic stars in Table 1, with
   NGC\,1624-2 explicitly labeled. The dashed blue lines mark
   approximate locations of the stellar surface and the iron-opacity
   bump. Figure adapted from \citet{Sundqvist13}.}
 \label{Fig:turb}
\end{figure} 

Using the method described in \citet{Sundqvist13}, the left panel of
Fig.~\ref{Fig:turb} shows fitted {\sc c\,iv} photospheric
line-profiles for three stars in the sample given in
Table~\ref{Tab:params}, namely HD\,191612, NGC\,1624-2, and the
non-magnetic comparison star HD\,36861. Since in the optical the
magnetic broadening due to the Zeeman effect is only $\sim$\,1-2\,km/s
per kG, both the magnetic (Table 1) and rotational contribution to the
line-broadening is negligible for HD\,191612, i.e., the total line
broadening may be quite unambiguously associated solely with
macroturbulence. Since the comparison star HD\,36861 reveals very
similar broad lines, this then suggests a common origin of the
observed macroturbulence in magnetic and non-magnetic O-stars.  By
contrast, the observed line in NGC\,1624-2 is qualitatively very
different, much narrower and with magnetic Zeeman splitting directly
visible (due to the very strong surface field, see
Table~\ref{Tab:params}). This indicates that the mechanism responsible
for the large macroturbulent velocities in the other stars is not
effective in NGC\,1624-2. The quantitative analysis by
\citet{Sundqvist13} results in $\theta_{\rm G} = \, \rm 2.2
\pm^{0.9}_{2.2}\,km/s$ for NGC\,1624-2. Such a very low (consistent
with zero) macroturbulent velocity is in stark contrast with the rest
of the sample, which displays $\theta_{\rm G} \approx \, \rm 20-65\,
km/s$ (Table 1). Thus, macroturbulence seems to behave similar in
non-magnetic and magnetic O-stars, except for in NGC\,1624-2 where it
is anomalously low or even completely absent.

\textbf{A simple model for magnetic inhibition of macroturbulence.} In
intermediate-mass Ap-stars, it is believed that the strong magnetic
field prohibits atmospheric motions between field lines and so
suppresses surface convection (e.g. \citealt{Balmforth01}; J.
Landstreet, priv. comm.). The critical parameter controlling the
competition between plasma and field in the atmosphere is the
so-called ``plasma $\beta$'', the ratio between gas pressure and
magnetic pressure, $\beta \equiv \frac{P_{\rm g}}{P_{\rm B}} =
\frac{P_{\rm g}}{B^2/(8\pi)}$ for magnetic field strength $B$. Let us
now thus assume the magnetic field stabilizes the atmosphere against
motions approximately down to the stellar layer at which $\beta
=1$. By adopting a very simple, classical gray model atmosphere, and
assuming a fossil field with no significant horizontal variations in
pressure and density between field lines, we obtain an analytic
expression for the temperature $T_0$ in the atmosphere at which $\beta
=1$ (see \citealt{Sundqvist13} for details),
\begin{equation}
  T_0 = T_{\rm eff} \left( \frac{3}{32 \pi} \frac{B^2 \kappa}{g} + 
  \frac{1}{2} \right)^{\frac{1}{4}}
  \approx 0.42 \, T_{\rm eff} B^{\frac{1}{2}} (\kappa/g)^{\frac{1}{4}},
  \label{Eq:T0}
\end{equation} 
where $B$ has units of Gauss and $\kappa \rm \ (cm^2/g)$ is a mean
mass absorption coefficient. The second expression here neglects the
1/2 within the parenthesis, and so implicitly assumes a field strength
significantly stronger than the $B \approx 400 \, (10^{-4}
g/\kappa)^{1/2}$ that yields $\beta =1$ at $T_0 = T_{\rm eff}$.

To estimate $T_0$ for the magnetic O-stars, the stellar parameters in
Table 1 are used together with the averaged surface field for $B$. For
simplicity, $\kappa = 1$ is further assumed for all stars; inspections
of Rosseland opacities in detailed {\sc fastwind} non-LTE model
atmospheres \citep{Puls05} show that for atmospheric layers with
$\tau_{\rm Ross} \ge 0.1$, such constant $\kappa \approx 1$ actually
is a quite good opacity-estimate for Galactic O-stars that are not too
evolved. The right panel of Fig.~\ref{Fig:turb} shows $T_0$ vs. the
$\theta_{\rm G}$ values given in Table 1 for the magnetic O stars. The
figure illustrates the influence of the magnetic field reaches down to
much deeper layers in NGC\,1624-2 than in any other star. This
suggests that the physical mechanism causing the large macroturbulence
in O-stars likely originates in stellar layers between 100\,kK and
200\,kK, consistent with a physical origin in the iron-peak opacity
zone located roughly at $T \approx 160$\,kK. Since the increased
opacity in this sub-surface zone is believed to trigger extensive
convective motions \citep[e.g.,][]{Cantiello09}, this makes the
analogy with suppression of surface convection in magnetic AP-stars
quite appealing.

\section{Summary and conclusions} 

Sect.~3 in this paper shows that the presence of significant
macroturbulence can result in severe overestimates of $v \sin i$ (at
least for rather slow rotators) when applying standard spectroscopic
methods (see also \citealt{SimonDiaz14, Aerts14}). This may have
important consequences, e.g., for determining the statistical
distribution of rotation rates for populations of massive stars, and
for the use of such rates in stellar evolution models (e.g.,
Ram{\'i}rez-Agudelo, this Volume).

The key for obtaining better constrained values of $v \sin i$ is a
more robust description of the so-called macroturbulent
line-broadening. Following \citet{Sundqvist13}, Sect. 4 places first
empirical constraints on the formation depth of such macroturbulence,
locating it to the region around the iron opacity-bump at $T \approx
160$\,kK. An attractive scenario then is that the responsible physical
mechanism is related to the convection believed to occur in this
region \citep[e.g.,][]{Cantiello09}, perhaps via stellar pulsations
excited by the convective motions \citep{Aerts09, Shiode13}.

\acknowledgements{JOS gratefully acknowledges support by the German
  DFG, under grant PU117/8-1.}

\bibliographystyle{iau307}
\bibliography{MyBiblio}

\begin{discussion}

\discuss{A. Herrero}{I agree that we badly need a better description
  of the broadening we observe in O-stars. Conserning the accuracy of
  classical methods (FT, GOF), in the recent paper by Simon-Diaz and
  myself we show that if there are other broadening mechanisms than $v
  \sin i$ and macroturbulence, we may obtain too large $v \sin i$
  values. On the other hand, in $\theta^1$ Ori C we get $v \sin i$
  values that agree with the rotation period derived from
  spectroscopic variations and B-field inclinations.}
\discuss{Sundqvist}{Yes, I am aware that in \citet{SimonDiaz14} you
  show that $v \sin i$ may be overestimated in the presence of large
  \textit{micro}-turbulence. I was not aware, however, that you
  obtained good agreement for $\theta^1$ Ori C. Note that I did not
  include this star here, since its rotation period 15 days actually
  implies a ``non-zero'' rotation speed. As such, the exact value of
  $v \sin i$ then depends on the uncertain stellar radius. But we
  should definitely investigate this further.}.

\discuss{S. Ibadov}{What can you say about generation of spots on
  massive stars surfaces, like sun-spots?}  \discuss{Sundqvist}{Note
  first that the magnetic fields I have been discussing here are
  large-scale, organized fields, with a dominant dipolar component and
  presumably of fossil origin. These fields are quite different from
  the complex, dynamo-generated fields in the Sun and other cool
  stars. That said, there have been some investigations regarding how
  a hypothetical magnetic field generated in the near-surface
  convection zone of massive stars could give rise to spots on the
  surface (e.g. Cantiello \& Braithwaite 2011). Such spots, however,
  would be \textit{hot and bright}, since the energy near the surface
  of massive stars is transported by radiation.}

\discuss{A. Lobel}{An important spectroscopic characteristic of yellow
  hypergiants ($T_{\rm eff} < 10\,kK$) are very broad photospheric
  absorption lines. They are slow rotators with large supersonic
  macroturbulence. Would you attribute its physical origin to g-modes
  in these cool massive stars as well?}  \discuss{Sundqvist}{That is
  difficult to say. The situation definitely seems reminiscent of that
  in blue supergiants, in which g-modes may indeed be the physical
  origin \citep[e.g.,][]{Aerts09}. But without looking further into
  the situation, I unfortunately cannot say much more than that at the
  moment.}

\discuss{C. Aerts}{Remark: We are including velocity fields due to 2-D
  (ideally in the future in 3-D) hydro simulations, and we do get
  broadened wings as suggested observationally. This is probably best
  explained with pulsations in gravity modes, as a ``natural''
  explanation for un-evolved B stars near the main sequence.}

\end{discussion}

\end{document}